# Polymer-based black phosphorus (bP) hybrid materials by *in situ* radical polymerization: an effective tool to exfoliate bP and stabilize bP nanoflakes.

Elisa Passaglia[a]*, Francesca Cicogna[a], Federica Costantino[a], Serena Coiai[a], Stefano Legnaioli[a], Giulia Lorenzetti[a], Silvia Borsacchi[a], Marco Geppi[a,b], Francesca Telesio[c], Stefan Heun[c], Andrea Ienco[d], Manuel Serrano-Ruiz[d], Maurizio Peruzzini[d]

[a]Istituto di Chimica dei Composti Organometallici (CNR-ICCOM), SS Pisa, Via Moruzzi 1, 56124 Pisa, Italy. E-mail: passaglia@pi.iccom.cnr.it

[b]Dipartimento di Chimica e Chimica Industriale (DCCI), via Moruzzi 13, 56121 Pisa, Italy

[c]NEST, Istituto Nanoscienze-CNR and Scuola Normale Superiore Piazza San Silvestro 12, 56127 Pisa, Italy

[d]Istituto di Chimica dei Composti Organometallici (CNR-ICCOM), Via Madonna del Piano 10, 50019 Sesto Fiorentino, Italy

**Abstract**

Black phosphorus (bP) has been recently investigated for next generation nanoelectronic multifunctional devices. However, the intrinsic instability of exfoliated bP (the bP nanoflakes) towards both moisture and air has so far overshadowed its practical implementation. In order to contribute to fill this gap, we report here the preparation of new hybrid polymer-based materials where bP nanoflakes exhibit a significantly improved stability. The new materials have been prepared by different synthetic paths including: i) the mixing of conventionally liquid-phase exfoliated bP (in DMSO) with PMMA solution; ii) the direct exfoliation of bP in a polymeric solution; iii) the *in situ* radical polymerization after exfoliating bP in the liquid monomer (methyl methacrylate, MMA). This last methodology concerns the preparation of stable suspensions of bPn-MMA by sonication-assisted liquid phase exfoliation (LPE) of bP in the presence of MMA followed by radical polymerization. The hybrids characteristics have been compared in order to evaluate the bP dispersion and the effectiveness of the bPn interfacial interactions with polymer chains aimed at their long-term environmental stabilization. The passivation of bPn results particularly effective when the hybrid material is prepared by *in situ* polymerization. By using this synthetic methodology, the nanoflakes, even if with a gradient of dispersion (size of aggregates), preserve their chemical structure from oxidation (as proved by both Raman and $^{31}$P-Solid State NMR studies) and are particularly stable to air and UV light exposure. The feasibility of this approach, capable of efficiently exfoliating bP while protecting the bPn, has been then verified by using different vinyl monomers (styrene and N-vinylpyrrolidone), thus obtaining hybrids where the nanoflakes are embedded in polymer matrices with a variety of intriguing thermal, mechanical and solubility characteristics.

**Introduction**

Black phosphorus (bP) is nowadays one of most studied 2D layered systems [1]. With its capability to form 2D structures similar to graphene, but with electronic properties potentially better suited to transistors or solar cells, bP and its exfoliated derivatives, phosphorene (a single layer of bP) or bP nanoflakes (bPn), have captured the attention of researchers around the world. Condensed matter physicists, chemists, semiconductor device engineers, and material scientists are in-depth studying the possible applications of bP and bPn in different fields[2,3]. Similarly to graphite and transition metal dichalcogenides (TMDCs), bP has a layered structure but a unique puckered single-layer geometry responsible for its interesting properties. The bPn have been reported to exhibit a high mobility of 1000 cm$^2$ V$^{-1}$ s$^{-1}$ for a sample of thickness 5 nm with high current ON/OFF ratio of 10$^5$ [4,5]. In addition due to the characteristic P atom arrangement, the carrier mobility is anisotropic in the plane and the direct electronic band gap depends on flake thickness[4-6]. In particular, bP is a p-type semiconductor which possesses a direct band gap of 0.3 eV; more interesting the bPn has again a direct band gap whichincreases up to approximately 2eV for the monolayer. These features make bP a promising material for novel applications in nanoelectronic and nanophotonic devices which cover the entire range of the visible spectrum.[3]. For example, exfoliated bPn (mechanically or via laser irradiation technique)[7,8] has been used in field emitter devices for the development of practical electron sources. Other interesting applications concern the use of exfoliated bP as humidity sensors [9,10] with performance depending on the thickness/number of nanosheets, and competitive with TMDCs [11,12]

However, the literature concerning the bP and bPn properties, their reactivity (including also the possibility of surface functionalization or decoring), and their stability in different environments, points out that the combination of air (oxygen), water (humidity) and light (UV radiations) causes the easy degradation of these materials[13-16]. It is also reported that the degradation is faster with decreasing layer number and thus flake thickness (on going from bulky bP to single layer bPn)[17]. This is for the moment one of the main drawbacks limiting the use of this material. Exfoliated bP has therefore to be stabilized to prevent degradation. This has been done through suitable coatings such as layered materials (hexagonal Boron Nitride)[18], oxides (Al$_2$O$_3$)[19], or polymers. Among these, polymer coating is the easiest, with particular reference to poly(methyl methacrylate) (PMMA) which is known to efficiently preserve mechanically exfoliated bP flakes[7]. Besides, mechanical exfoliation of bP is not a scalable technology which limits its use in applications.

Notably, polymers have been recognized as favorable in breaking down the strong interlayer interactions among nanostructured layered materials[20] to design hybrids with performance suitable for optoelectronics or nonlinear optical (NLO) devices, chemiresistors, temperature and deformation sensors[21]. Depending on the possibility to establish specific interactions, the agglomeration of thin bP layered flakes, once generated by liquid-phase exfoliation (LPE) technology, can be avoided by embedding them with polymer chains, like in the case of polystyrene[22]. In addition the final polymer nanocomposites can preserve the structure and the properties of individual components, realize synergistic effects between different substrates, endow new properties and develop devices for different applications; PEDOT:PSS, poly-L-lysine, polyaniline, polycarbonate have been used to design semiconductors, sensing platforms for medical detection, pseudocapacitors, and pulsed fibre lasers[23-26] which are generally obtained by covering the bPn surfaces or by mixing their suspension with the polymers.

By considering this scenario, the present work proposes, for the first time, a study focused on the preparation of stabilized bPn embedded in polymers through hybrids preparation able to maintain or even to improve the structural properties of bP. In place of the previously reported exfoliation methods that use a mechanical tool (scotch tape) or LPE with non-friendly solvents, that are difficult to remove from the devices (as for example DMSO), here, for the first time, bP has been exfoliated in the PMMA matrix by using strategies aimed at improving both the exfoliation level (phosphorene or bPn being the target) and the structural and morphological stability of the exfoliated material. In particular, the bP has been directly exfoliated in the liquid vinyl monomer, the methyl methacrylate (MMA), without solvents, and after the addition of a radical initiator, the hybrid material has been obtained by *in situ* radical polymerization. For comparison purposes PMMA-based composites have been also prepared by the direct bP exfoliation in polymer solution or by starting from bPn previously exfoliated by LPE with DMSO and successively dispersed in the PMMA matrix.

These new materials have been characterized by combining different techniques: Dynamic Light Scattering (DLS), Fourier Transform Infrared (FTIR) and Raman spectroscopies, X-ray Diffraction (XRD), $^{31}$P-Solid State NMR (SSNMR), Atomic Force Microscopy (AFM), Size Exclusion Chromatography (SEC), Differential Scanning Calorimetry (DSC), Thermal Gravimetric Analysis (TGA) have been used to in-depth examine the structural features, the morphology and the thermal properties of components as soon as the hybrids have been obtained and also after several months. The collected results gather information about the stability of the chemical structure of bPn in the polymer matrix over time and after UV light irradiation in air. The *in situ* radical polymerization has been thus used to produce hybrids starting from different vinyl monomers, styrene (Sty) and N-vinylpyrrolidone (NVP) to prove the feasibility and the significance of the methodology here developed.

**Experimental part**

*Materials*

All the materials (polymers and reagents) and the solvents below were used as received without further purification: methyl methacrylate (MMA) 99% from Sigma Aldrich, d = 0.963 mg/mL MW: 100.12 g/mol; 1-vinyl-2-pyrrolidone (NVP) ≥ 99% from Sigma-Aldrich, d = 1.043 mg/mL MW: 111.14 g/mol; styrene (Sty) ≥ 99% from Sigma Aldrich d = 0.906 mg/mL.

Poly(methyl methacrylate) (PMMA) from Sigma Aldrich MW: 120,000 D; polystyrene (PS) from Repsol MW: 164,500 D; poly(vinyl pyrrolidone) (PVP) from Sigma Aldrich MW: 29,000 D.

2,2′-Azobis(2-methylpropionitrile) (AIBN) 98% from Sigma Aldrich MW: 164.21 g/mol; dimethyl sulfoxide (DMSO) ACS reagent ≥99.5% from Sigma Aldrich MW: 78.13 g/mol; chloroform ACS reagent ≥ 99,8 % from Sigma Aldrich MW 119.38 g/mol; acetone ACS reagent ≥ 99,5% from Sigma Aldrich, MW: 58.08 g/mol; methanol ACS reagent 99.8% from Sigma Aldrich MW: 32.04 g/mol; diethyl ether ACS reagent ≥99.8% form Carlo Erba (CAS: 60-29-7) MW: 74.12 g/mol; propanol ACS reagent ≥99.5% from Sigma Aldrich (CAS: 67-63-0) MW: 60.10 g/mol; *n*-heptane ACS reagent ≥98.5% from Sigma Aldrich (MW: 100.2 g/mol; anisole 99% from Sigma Aldrich MW: 108.14g/mol

*Instruments and Characterization*

The micro-Raman analysis was performed using a Renishaw micro-Raman inVia instrument equipped with a 1800 grooves/mm diffraction grating, a CCD detector and a 50X magnifying lens. The instrument has a Nd:YAG laser source at λ = 532 nm wavelength. The samples were analyzed as polymer films or powder (bP); the measurements as well as the imaging were obtained on different portions of each specimen and the power on the samples was about 1.5 mW.

X-ray Diffraction (XRD) patterns of hybrids and bP were acquired at room temperature with a PANalytical X'PERT PRO diffractometer, employing Cu Kα radiation (λ = 1.54187 Å) and a parabolic MPD-mirror for Cu radiation. The diagrams were acquired in a 2θ range between 5.0° and 80.0°, using a continuous scan mode with an acquisition step size of 0.0263° or 0.0131 ° and a counting time of 150 s.

Dynamic Light Scattering (DLS) analyses were carried out at room temperature by using the Malvern Zetasizer nano instrument (model: ZEN1600) equipped with a HeNe laser (633 nm, 4 mW) and an avalanche photodiode detector with an angle of 173°. The DLS data were processed and analyzed with Dispersion Technology Software (Malvern Instruments).

Fourier Transform Infrared (FTIR) and Attenuated Total Reflectance (ATR-FTIR) spectra were recorded at room temperature with a Perkin-Elmer Two Spectrometer equipped with an ATR accessory with diamond crystal. The spectra were generally acquired between 4000 and 400 cm$^{-1}$ with a resolution of 4 cm$^{-1}$ using 16 scans.

Number average molecular weight ($\overline{M_n}$) and weight average molecular weight ($\overline{M_w}$) as well as dispersity (Đ) were determined using Size Exclusion Chromatography (SEC), Agilent Technologies 1200 Series. The instrument was equipped with an Agilent degasser, an isocratic HPLC pump, an Agilent refractive index (RI) detector, and two PLgel 5 μm MiniMIX-D columns conditioned at 35 °C. Chloroform (CHCl$_3$) was used as the mobile phase at a flow rate of 0.3 mL min$^{-1}$. The system was calibrated with polystyrene standards in a range from 500 to $3 \times 10^5$ g mol$^{-1}$. Samples were dissolved in CHCl$_3$ (2 mg mL$^{-1}$) and filtered through a 0.20 micron syringe filter before analysis (twice in the case of hybrids). Number average molecular weight ($\overline{M_n}$) and weight average molecular weight ($\overline{M_w}$) were calculated using the Agilent ChemStation software.

Thermal Gravimetric Analyses (TGA) were carried out with a Seiko EXSTAR 7200 TGA/DTA by introducing about 5-8 mg of sample in an alumina sample pan of 70 μL. In a typical experiment, run was carried out at a standard rate of 10°C/min from 30°C to 700 °C under nitrogen flow. T$_{onset}$ and T$_{infl}$ were determined by analyzing the TG curve (as the temperature of intercept of tangents before and after the degradation step) and DTG curve (as the maximum of the peak), respectively.

The glass transition temperature (Tg) of hybrids was determined by Differential Scanning Calorimetry (DSC) using a PerkinElmer DSC4000 equipped with intracooler and interfaced with Pyris software (version 9.0.2). The range of temperatures investigated was 40–180 °C. Thermal scans were carried out on 5–10 mg samples in aluminum pans under nitrogen atmosphere. The instrument was calibrated by the standards In (Tm = 156.6 °C, ΔHm = 28.5 J/g) and Pb (Tm= 327.5°C, ΔHm = 23.03 J/g ).

Atomic Force Microscopy (AFM) measurements were performed with a Bruker Dimension Icon AFM, in pick force mode. Data analysis was performed by WSxM software[27].

$^{31}$P Solid State NMR (SSNMR) experiments were carried out with a Varian InfinityPlus spectrometer working at Larmor frequencies of 400.34 Mz and 162.07 MHz for $^1$H and $^{31}$P nuclei, respectively. Spectra were acquired using a 3.2 mm probehead, exploiting the Direct Excitation (DE) pulse sequence, under High Power Decoupling from $^1$H nuclei, using a recycle delay of 120 s and accumulating a number of transients between 100 and 3000. All the experiments were carried out under Magic Angle Spinning (MAS), with a frequency of 10 KHz, using air as spinning gas, and at a temperature of 20 °C. $^{31}$P chemical shift scale was referred to the signal of $H_3PO_4$ (85%) at 0 ppm.

*bP and bPn suspensions preparation and characterization*

Black phosphorus (bP) and phosphorene (bPn) suspension in DMSO (DMSO_bPn, $r_H$ = 500 ± 23nm) were prepared as previously described[16] and TGA, XRD and Raman spectroscopy results agreed with reported data [22]. A compendium concerning the novel Raman analyses of bP is here reported in ESI (figure S1): the un-exfoliated bP crystals show the three characteristic peaks of modes $A_g^1$, $B_{2g}$, $A_g^2$ with a shift in peak positions depending on the number of layers. In agreement with data already reported in the literature[14,16,17,28], the observation of sharper peaks, which were weakly shifted towards higher wavenumbers, was taken as evidence for crystalline, thin bP sheets.

MMA_bPn, NVP_bPn, Sty_bPn suspensions were obtained by LPE in the presence of the sole monomer. In a typical procedure ~5 mg of bP carefully crashed in a mortar, were put in a test tube and then a weighted quantity of MMA, Sty or NVP was added (Table 1). The monomer bP suspension was sonicated for 90 min by using a Hielscher Ultrasonic Processor (UP220St) instrument, equipped with Sonotrode (diameter: 2 mm; 26±kHz). The amplitude of ultrasound wave was maintained constant at 50% with P = 7 Watt. In all cases an ice bath was used to avoid overheating of the system. The final MMA_bPn, NVP_bPn, Sty_bPn suspensions were then insufflated with $N_2$ for 15 min. All the suspensions were analyzed by DLS, showing $r_H$ values really close to that of DMSO_bPn. For example the MMA_bPn suspension (having about 1% of bP content) was characterized by $r_H$=512±58 nm.

*Hybrid materials preparation*

Three different methodologies were employed (see Figure S2 in ESI):

METHOD A: dispersion of DMSO_bPn suspension in PMMA solution.

Into a 100 mL two-necked round bottom flask, equipped with a magnetic stirrer and previously degassed, back-filled three times with nitrogen and then left under nitrogen, 25 mL of $CHCl_3$ and 0.523 g of PMMA (commercial product) were loaded. The solution was magnetically stirred for 10 min in a continuous stream of $N_2$ until the polymer was completely dissolved. Under a $N_2$ current, the DMSO_bPn suspension ($r_H$= 500±23 nm) was added dropwise. The mixture was left stirring under $N_2$ for 15 min and then the mixing stopped. A DLS measurement of polymeric suspension provided a value of $r_H$= 894.3±11.5 nm. The flask content, a yellow/brown solution, was precipitated (dropwise) into 400 mL of MeOH. The polymer was then filtered, and dried under vacuum till constant weight (0.430 g). By considering that the DMSO_bPn suspension was prepared by using 5mg of bP, the content of phosphorus derivative in the composite (entry PMMA_bP_A) was estimated as 1.0 % wt on the basis of starting amount.

<u>METHOD B</u>: LPE of bP in PMMA solution.

5 mg of bP flakes were put into a test tube; then a polymer solution containing 0.513 g of PMMA (commercial product) in 30 mL of mixture acetone/DMSO 2/1 was added to the powder. The ultrasonication process was carried out for 3 hs, by varying the amplitude of the ultrasound wave between 50 and 100%, with a power of 5-9 Watt. A DLS measurement of polymeric suspension provided a value of $r_H$=427.3 ± 96.4 nm. The collected dispersion with yellow-brown color was co-precipitated in 400 mL of MeOH. The solid fraction was then recovered via filtration and dried under vacuum until constant weight. The content of phosphorus derivatives in the composite (entry PMMA_bP_B) was estimated as 1.0 % wt on the basis of starting amount.

The same treatment (solubilization and sonication) was applied to PMMA (commercial product) without adding the bP to recover a blank sample (entry PMMA_B_blank) used for comparison purpose.

<u>METHOD C</u>: *in situ* radical polymerization

A Schlenk tube (10 mL) equipped with a magnetic stirrer and previously degassed, back-filled three times with nitrogen and then left under nitrogen, was loaded with the MMA_bPn or Sty_bPn or NVP_bPn suspensions (previously prepared). A weighted amount of AIBN (2 wt.% with respect to the monomer) was added and the tube was placed in an oil bath (temperature and time of polymerization depending of vinyl monomer used, as summarized in Table 1). The product was then dissolved in $CHCl_3$ and precipitated in an appropriate solvent to remove un-reacted monomer and polymerization by-products: MeOH was used for PMMA and PS, while $Et_2O$ for PNVP. After recovering via filtration, the resulting powder was dried under vacuum until constant weight. The amount of phosphorus derivatives in each composite was determined on the basis of the polymerization yield (ranging from 85 to 60 %) and its starting amount.

With the same experimental conditions MMA, Sty and NVP were polymerized to provide comparative samples obtained without bP (entries: PMMA_C_blank; PS_C_blank; PNVP_C_blank).

**Table 1**. *In situ* radical polymerization runs: experimental conditions and final composition of composites.

| Entry | Monomer (g) | | bP (g) | T (°C) | Time (min) | Final content of P (wt.%)* |
|---|---|---|---|---|---|---|
| **PMMA_bP_C** | MMA | (0.94) | 0.0055 | 70 | 180 | 0.8 |
| **PS_bP_C** | Sty | (0.91) | 0.0053 | 80 | 180 | 0.8 |
| **PS_bP_C2** | Sty | (3.62) | 0.0055 | 80 | 180 | 0.2 |
| **PNVP_bP_C** | NVP | (1.04) | 0.0050 | 75 | 150 | 0.7 |
| **PNVP_bP_C2** | NVP | (4.16) | 0.0063 | 75 | 150 | 0.3 |

* by supposing that the whole phosphorus amount was maintained in the composite

A schematic representation summarizing the different procedures is reported in Fig. S2.

In addition a physical mixture between bP and PMMA (1wt %) was prepared to be analysed by $^{31}$P-SSNMR and XRD.

All the samples were analyzed by SEC, DSC, TGA, and PMMA-based hybrids by $^{31}$P-SSNMR. In addition all the composites were even molded into films, by using a press Carver bench model 4386 (T = 180 °C, 10-20 Kg/cm$^2$), with constant and uniform thickness= 40-90 µm to be analyzed by XRD, Raman and FTIR-ATR. Photodegradation of PMMA_bP_C and PMMA_C_blank was studied using a UV-Vis camera (UvaCube400, 400W, Hoenle) equipped with a Hg lamp (high pressure Mercury Lamp with a power of 400W: emittance$_{230-285}$ = 15 mW/cm$^2$; emittance$_{330-400}$ = 11 mW/cm$^2$; emittance$_{380-500}$ = 35 mW/cm$^2$). The samples were irradiated for 250 min from one side. In addition, a solution of the sample PMMA_bP_C in anisole (23 mg/2mL) was spin coated at 4000 revolutions per minute (rpm) for 1 minute, after an acceleration step at 500 rpm for 5 sec, and analyzed by AFM. The samples PNVP_bP_C and PS_bP_C were solubilized by water and anisole, respectively, and films provided by solution casting were analyzed by Raman.

**Results and Discussion**

**PMMA-based hybrid materials**

The different methodologies used to prepare the hybrid material (schematized in Fig S2) can be summarized as follows: i) embedding of the already exfoliated bP (by conventional LPE) in PMMA; ii) exfoliation of bP by PMMA solution; iii) LPE of bP by monomer (MMA, Sty, NVP) followed by *in situ* radical polymerization. These synthetic approaches were designed and realized with the dual objective of achieving a good dispersion of bP, *i.e.* obtaining thin flakes or few layers flakes, and, at the same time, protecting them from the degradation that is known to occur when bP nanoflakes are exposed to air and light. The main target of the study was the achievement of a processable, soluble and stable hybrid material that contains thin bP flakes, or a few layer flakes, whose structure can be preserved for long time.

The structural and thermal properties of PMMA matrix in the hybrids were investigated by FTIR, Raman, SEC, TGA and DSC (Figure 1 and Table 2). The FTIR and Raman spectra of the samples showed all the characteristic absorption bands of PMMA matrix, whose attributions are reported in ESI_Table S1[29]. No differences in the FTIR spectra of composites with respect to the spectrum of commercial sample or blank experiment was highlighted (see Figure 1A, as example) suggesting that the different synthetic paths and the presence of bP or bPn did not cause significant variations in the chemical structure of PMMA. Notably, FTIR spectra did not show any additional absorption bands due to bP. Raman spectra (Figure 1B) confirmed the presence of PMMA showing all the main absorptions peaks of the matrix (Table S1), which were not changed by the presence of bP derivatives. In addition, these spectra showed distinct signals of bPn (between 300 and 500 cm$^{-1}$) whose intensities and Raman shifts depended on the methodology used for the preparation of the samples (as discussed later).

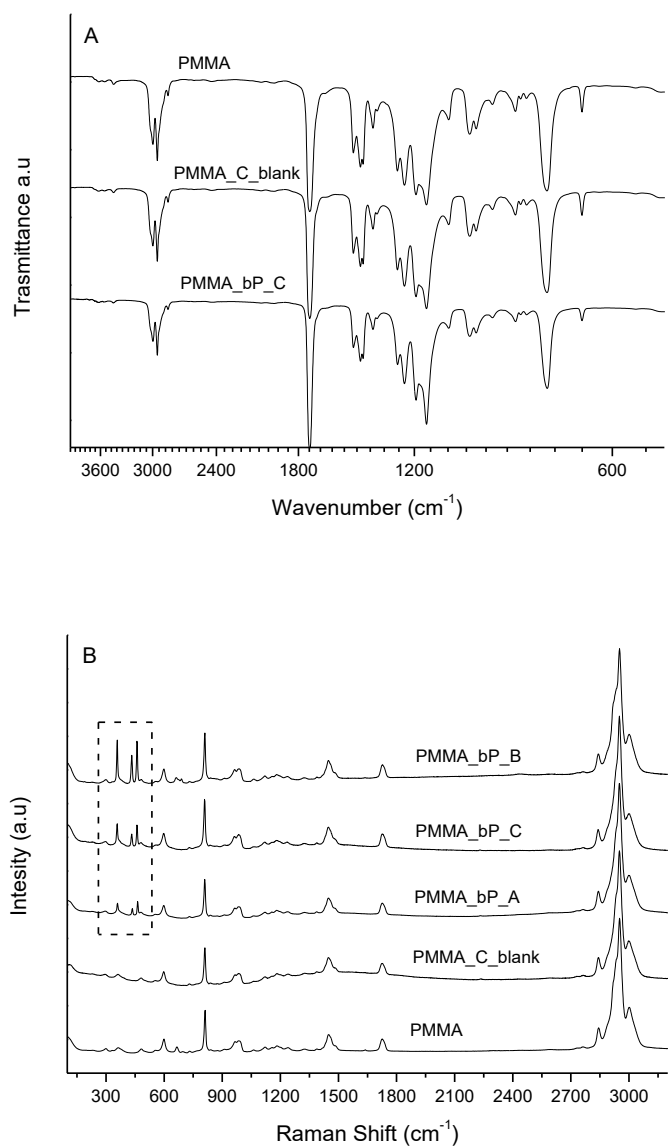

**Figure 1.** A: FTIR spectra of commercial PMMA, and of samples PMMA_C_blank and PMMA_bP_C. B: Raman spectra of matrices (PMMA and PMMA_C_blank) and of hybrids obtained by different methods (dotted box highlights the bP signals and confirms its presence in all hybrids).

The SEC analysis (Table 2) showed that the PMMA phase of the hybrids prepared by METHOD A and METHOD B has the same or similar $\overline{M_n}$ with respect to its reference (i.e. PMMA and PMMA_B_blank, respectively) with only a weak decrease of $\overline{M_w}$ for samples obtained by METHOD B. This result indicates a possible degradation of polymer chains induced by the prolonged sonication treatment. Conversely, the sample PMMA_bP_C, prepared by *in situ* radical polymerization, was characterized by a remarkably higher value of both the average molecular weights. This experimental evidence can be explained by the hindrance in movement of the growing macroradicals which inhibits the termination reactions and thus increases the length of polymer chains[30]. This confinement effect confirms that the growing of PMMA macromolecules occurred near or onto the bPn surfaces or possibly between the layers of bP and promoted an effective embedding of flakes with the polymer chains.

**Table 2:** Molecular weight evolution and thermal features of PMMA-based samples

| Sample | $\overline{M_n}$ (D) | $\overline{M_w}$ (D) | $T_g$ (°C) | $T_{onset}$(°C)* | $T_{infl}$(°C)$^§$ |
|---|---|---|---|---|---|
| PMMA | 52,000 | 101,000 | 105.0$^\#$ | 264 | 290-387 |
| PMMA_bP_A | 56,000 | 97,000 | 115.6 | 279 | 294-394 |
| PMMA_B_blank | 57,000 | 90,000 | 108.7 | 267 | 285-390 |
| PMMA_bP_B | 49,000 | 80,000 | 115.1 | 280 | 294-395 |
| PMMA_C_blank | 45,000 | 103,000 | 120.6 | 272 | 287-381 |
| PMMA_bP_C | 58,000 | 198,000 | 121.0 | 269 | 293-372 |

* intercept of tangents before and after degradation step; $^§$ from DTG curves as maximum of peak; $^\#$ from technical sheet

AFM analysis of films obtained by spin coating of PMMA_bP_C anisole solution corroborated the evidence of strong interactions and entanglement between polymer chains and bPn. PMMA fractions densely aggregated and formed a net around smaller particles that showed the characteristic Raman peaks of bP. An example of these hybrid PMMA/bP aggregates is reported in Figure 2; the "*plateau*" area visible in Fig. 2 (a) is 4 nm higher than the surrounding PMMA thin film and 4-5 µm wide (Fig. 2(d)), while the bPn is inhomogeneous and up to 200 nm high, as shown in Fig. 2(e). Zooming in and rescaling the image to properly see the flake (Fig. 2(b)), we can observe that the 1µm bPn is indeed an aggregate of smaller structures. This inhomogeneity, as well as the height difference between the bP aggregate and the plateau, is even more evident from the 3D visualization (Fig. 2(c)). Therefore, sample PMMA_bP_C actually contains a portion of PMMA strongly interacting with the bP flakes. It presumably grew up from the layers within the same flake and is characterized by higher molecular weight. Even if only a few literature examples of bP covalent functionalization are reported, we cannot completely exclude that bP sites (presumably the P apical atoms) are involved in MMA polymerization, by generating P-C bonds[31]. The reactivity of elemental white phosphorus with carbon-centered radicals is well-known[32] and its use as alkyl radical trap is well-documented[33]; in addition, some weak hypothesis about the radical reaction of bP with aryl radicals (derived from diazonium compounds) was recently discussed as capable of generating P-C covalent bonds[34,35]

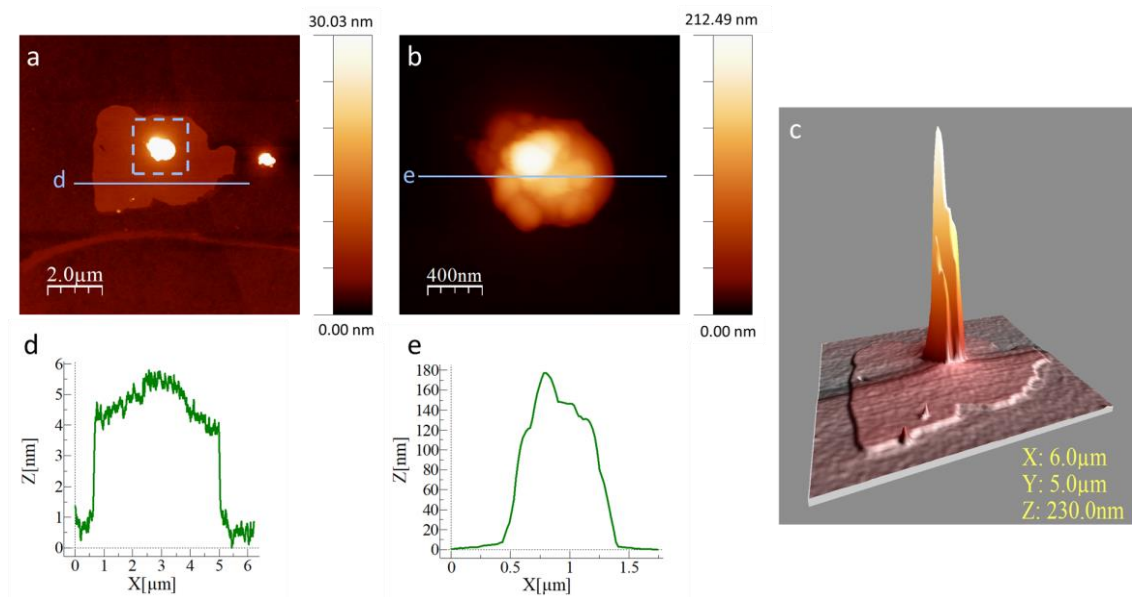

**Figure 2:** AFM analysis of a film obtained by spin coating of PMMA_bP_C anisole solution. The film has a thickness of approx. 20 nm, as measured by stylus profilometer. Panel (a) shows a small aggregate of bP surrounded by a several micron-wide "plateau". This "plateau" is composed of densely packed polymer chains, aggregated around the bP structure. (b) Zoom-in taken in the region indicated by the square box in (a). It displays the bP structure, which appears as a composite of individual bP flakes. (c) 3D representation of the region of interest of (a), which allows to appreciate the difference in height between the "plateau" and the bP. (d) Cross section of the "plateau" taken along the line shown in (a). The "plateau" is ~4 nm higher than the surrounding area. (e) Cross section of the bP aggregate, along the line shown in (b), displays the height of the bP aggregate, up to approximately 200 nm, and its inhomogeneity.

All hybrids showed an increase of $T_g$ value with respect to their blank experiment or reference (Table 2) suggesting a reinforcing effect due to nanofiller addition; however, this increment occurred with a really low extent for the run obtained by *in situ* polymerization (METHOD C) suggesting for this sample a finer and homogeneous dispersion of flakes. The same trend was observed for the TGA results. Notably, the composites provided by METHODS A and B showed a certain improvement in their thermal stability with respect to both the onset and inflection temperatures, in agreement with results already reported[16]. PMMA_bP_C has a thermal behaviour similar to that of blank sample, suggesting the formation of an interpenetrated phase in which the two components (polymer and filler) are really entangled at molecular level without distinguishable effects regards to bulk thermal properties (the TGA curves are reported in Figure S3)

The optical microscopy coupled with Raman was used to investigate the morphology of the samples; the imaging of portions of each specimen showed a different flakes distribution, depending on the preparation method (Figure 3). PMMA_bP_A showed a good homogeneous distribution of tiny particles (below 1 μm) resembling the bPn and a really small amount of larger aggregates (see an example on the right of Figure 3 A). Instead, the PMMA_bP_B sample seemed to be characterized by the presence of very large inclusions (Figure 3 B), even if the $r_H$ measured of the polymeric suspension is comparable with that of DMSO_bPn (see experimental part). Notably, a finer dispersion of bP was achieved in the case of PMMA_bP_C since tiny particles and flakes, homogeneously distributed, were observed and only a small fraction of larger aggregates (of several μm) was found (see an example on the right of Figure 3 C).

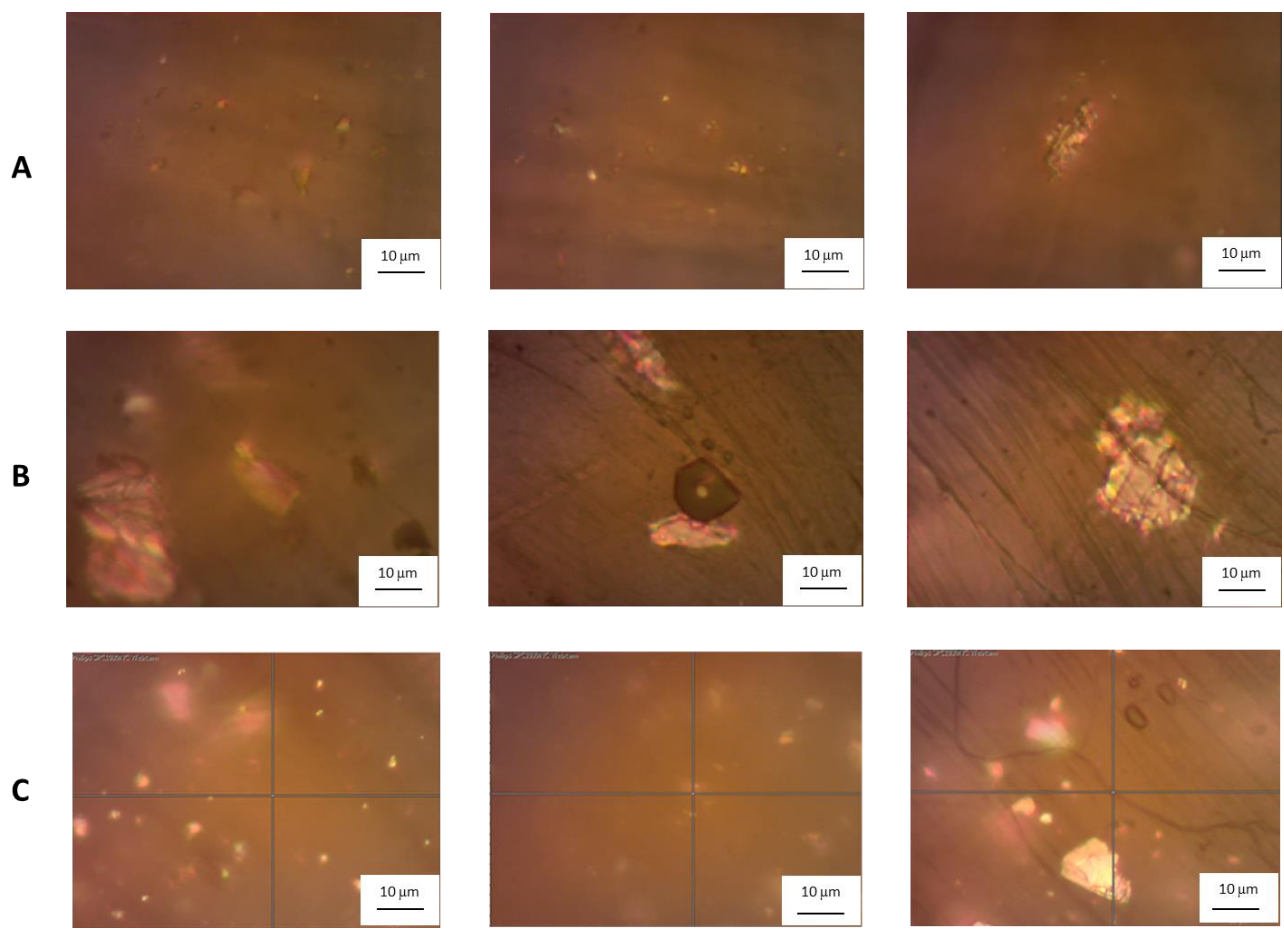

**Figure 3:** Magnified visual images of hybrids produced by methods A, B, and C, collected by optical microscope and showing the particles' distribution and their dimensions.

Raman spectra were collected in different portions of the samples to focus on the structure and distribution of flakes. A comparison of representative spectra collected for all the hybrids is reported in Figure S4. All the samples showed signals confirming the presence of bP, which suggest that most of the nanostructured material survived the methodology used for the hybrid synthesis, even if in the case of sample prepared by METHOD A not all the detected flakes exhibited the characteristic Raman peaks of bP structure. Similar behaviour was observed for sample PMMA_bP_B for which only the large particles showed very intense signals attributable to bP; while in the case of sample PMMA_bP_C almost all the observed flakes exhibited the $A_g^1$, $B_{2g}$, $A_g^2$ modes[14,16,17,28]. In addition, the spectrum of Raman active flakes of sample PMMA_bP_A showed a blue-shift in peak positions (Figure S4), suggesting the presence of thinner flakes[9, 28], in agreement with the morphological evidences.

Interestingly, from inspection of the structural features of flakes detected in samples provided by METHOD B and C, it was found that the intensity of bP peaks compared to those of polymer was different and depended on the spot investigated. PMMA_bP_B showed a remarkable decrease of bP signal intensity by moving from the particle to the apparently neat polymer; moreover, Raman spectra of larger particles (several microns) did not evidence the typical polymer signals, according to poor efficiency in the bP embedding in the polymer matrix (Figure 4 A, spectrum (a)). Instead, the spectra of PMMA_bP_C disclosed, in all the analysed portions, the vibration modes of both the components, bP and polymer, with comparable relative intensity (Figure 4 B). Smaller particles/aggregates or apparently neat polymer portions were characterized by bP signals shifted towards higher frequency. This evident blue-shift of the $A_g^2$ mode (suggested as the most sensitive indicator of layer number[9, 36]) and the decrease of intensity with respect to the reference PMMA band (see figure 4C), confirmed for this sample the presence of almost homogeneously distributed thinner flakes (or bPn), consistent with the observations concerning the morphological features of the different hybrids discussed above. We speculate that the growth of polymer chains near or between the nanolayers, which is typical of the *in-situ* polymerization technique, allows the moving away of bP layers and thus the obtainment of thinner flakes

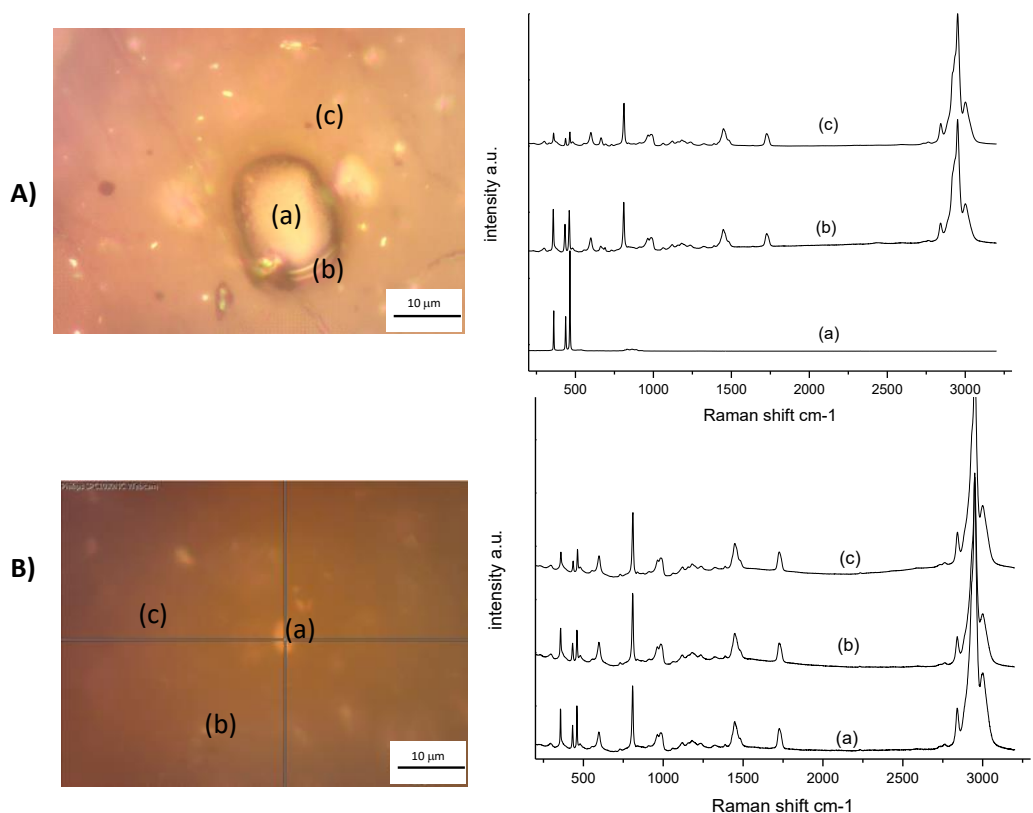

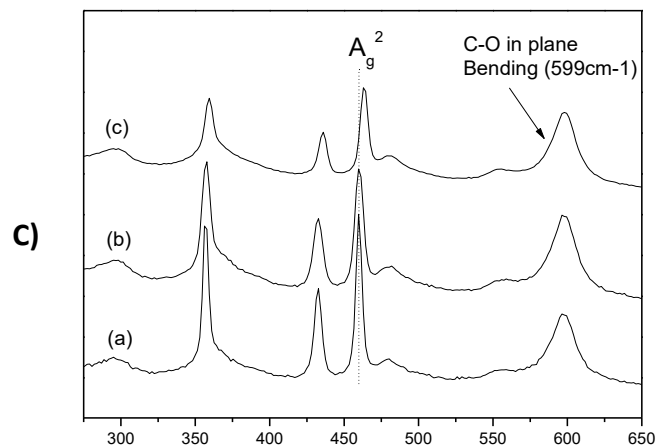

**Figure 4**: Representative images and Raman spectra collected in point indicated by letters (a), (b) and (c) of A): sample PMMA_bP_B and B): sample PMMA_bP_C; C): enlargement of Raman spectra in the region of bP modes for PMMA_bP_C sample (dotted line is guide for eyes).

Solid State Nuclear Magnetic Resonance (SSNMR) spectroscopy is at present one of the best techniques to characterize structural and dynamic properties of solid materials, over wide spatial and time ranges and independently of their amorphous or crystalline character[37].

$^{31}$P-MAS spectra (with or without $^{1}$H HPD) were recorded to characterize bPn after embedding in PMMA. In the literature only a few examples of $^{31}$P-MAS spectra of bP are reported [38-40] while, to the best of our knowledge, these are the first spectra of hybrids materials containing bP.

Figure 5 shows the $^{31}$P-MAS spectra of bP physically mixed with PMMA (see experimental part) used as reference sample and of PMMA hybrids, prepared by using the different methodologies.

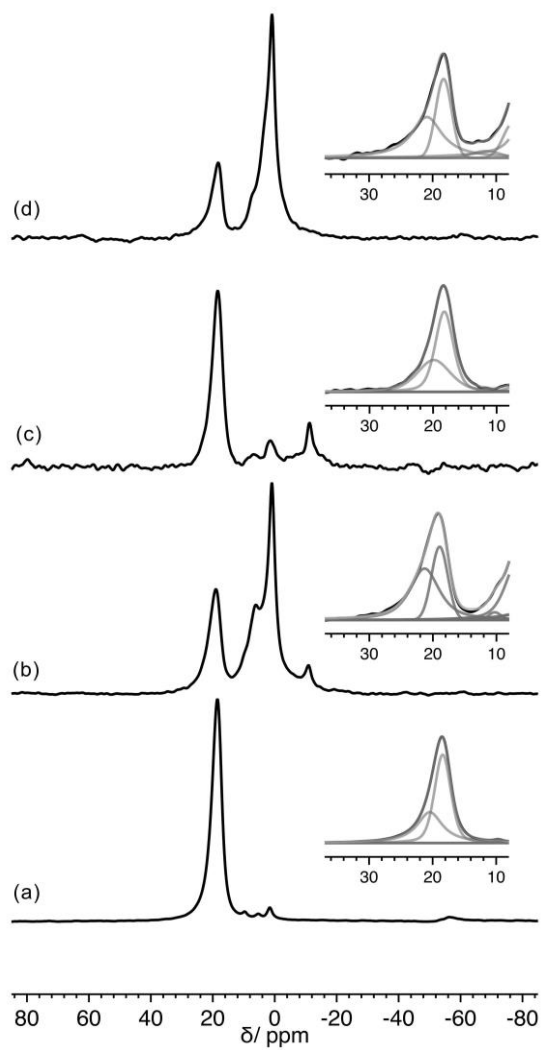

**Figure 5:** $^{31}$P-MAS NMR spectra of (a) physical mixture between PMMA and bP, (b) PMMA_bP_B, (c) PMMA_bP_C, and (d) PMMA_bP_A. Differently from spectra (b)-(d), spectrum (a) was recorded without HPD from $^1$H nuclei. The inset of each spectrum shows the fitting of the bP signal resonating at 18.5 ppm.

The spectrum of bP physically mixed with PMMA (Figure 5 (a)) showed a signal at 18.5 ppm, consistent with the few bP NMR spectral data reported in the literature[38-40]. Moreover, two small signals were present at 0.8 ppm (singlet) and 7.5 ppm (doublet, with a $J_{(P-H)}$ = 350 Hz), mainly ascribable to $H_3PO_4$ and $H_3PO_3$ species, respectively[40,41]. These signals indicate that some oxidation effects occurred, even if to a small extent (the total area of $H_3PO_4$ and $H_3PO_3$ signals account for 6% and 3% respectively, of the whole spectral area). The presence of oxidized species is likely due to the chemical adsorption of oxygen on the bP surface[42] which was not protected, forming aging products.

All the hybrids materials showed the signal of bP at *ca.* 18.5 ppm and several peaks in the 0÷15 ppm spectral region, ascribable to oxidation products, mainly $H_3PO_4$ and $H_3PO_3$, but possibly also other phosphates and oxidized species[43,44]. Moreover, in the spectra of PMMA_bP_B and PMMA_bP_C weak signals were present in the region -3÷-20 ppm, with a peak at -11 ppm, ascribable to pyrophosphate[41]. For samples PMMA_bP_A and PMMA_bP_B the intensity of the signals of the oxidized products was relatively high, about 75% and 71% of total spectral intensity, respectively, suggesting extensive degradation of bPn in the conditions used for the samples preparations (METHODS A and B see Fig. S2) and in agreement with the observation that not all bP flakes in these samples were Raman active. Instead, the spectrum of the sample prepared by *in situ* polymerization, METHOD C, (PMMA_bP_C) showed a higher intensity of the bP signal and a lower intensity of the signals due to the degradation products (about 68% and 32%, respectively). From these results it appears evident that METHOD C better preserved the bPn structure. Conversely, METHOD A involved the use of previously exfoliated bP that was co-precipitated in a solvent after mixing with the polymer solution. Thin flakes easily underwent degradation/oxidation during the work-up, particularly during the treatment with solvents upon prolonged sonication. Sample prepared by METHOD B was sonicated for longer time (3 h). Although this procedure was necessary to boost the bP exfoliation in the polymer solution, the fact that we could not operate under inert atmosphere, rendered this methodology the less suited to guarantee the complete bP structure preservation. Instead, the LPE in the MMA monomer carried out under milder conditions and the subsequent *in-situ* radical polymerization provided the best results in terms of bPn structure stability. In addition, the presence of weak signals in the region 10 ÷ -20 ppm could be ascribed to alkyl-phosphorus species originating from the reaction between organic radicals and bP or bPn.

Remarkably, in the spectrum of the physical mixture between PMMA and bP, the bP signal at 18.5 ppm shows an asymmetric shape, observed also in the already reported bP spectra. Indeed, by exploiting a spectral fitting procedure, the signal at 18.5 ppm could be deconvoluted in two peaks, the first with a chemical shift of 18-19 ppm (linewidth 450-550 Hz) and the second at 20-21 ppm (linewidth 800-1000 Hz; see inset in Figure 5). The intensity ratio between these two peaks was about 60:40. Approximately the same result was obtained for PMMA_bP_C (inset of Figure 5), suggesting that the exfoliation degree did not substantially affect the chemical shift and the shape of the $^{31}$P NMR signal.

The bP NMR signal of PMMA_bP_A and PMMA_bP_B appears even more asymmetric, as confirmed by a 40:60 intensity ratio between the peaks at about 18-19 ppm and 20-21 ppm, as determined from spectral fitting (insets of Figure 5). Considering that these two samples, even if exfoliated to a different extent, present a similarly high degree of oxidation, this result suggests that a large degradation could also affect the signal of non-oxidized phosphorus atoms, increasing the component at higher chemical shift.

X-ray diffraction analysis was used to characterize the crystalline forms of neat bP (mixed with PMMA) and after being dispersed in the hybrids. The typical XRD patterns collected at room temperature in the scanning range 5° < 2θ < 60° are reported in Figure 6. Broad bands at 2θ = 13.8°, 30° and 41.6° were observed for all the samples, confirming the amorphous nature of polymer[45]. The XRD pattern of the physical mixture PMMA/bB (Figure 6 (a)) having composition similar to that of the hybrids, showed the typical bP diffraction peaks (020) (040) and (060)[45], centred at 2θ: 16.90°, 34.19° and 52.34°, respectively. The same characteristic diffraction peaks were present in the XRD patterns of the hybrids. Moreover, it was evident that the intensity of the peaks associated to the crystalline fraction of bP along the *z* direction was different, suggesting a different degree of order in this direction, likely meaning that the average number of piled layers was not the same and depended on the kind of sample. In fact, the preparation methodologies are responsible of the content of bP able to preserve its structure and of the content of exfoliated bP (bPn) whose nanoflakes theoretically should not be ordered and piled[46].

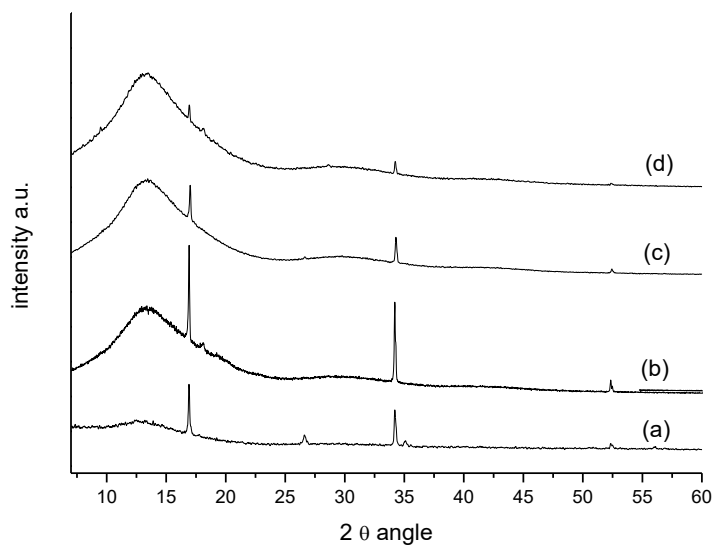

**Figure 6**: XRD patterns of physical mixture between (a) PMMA and bP, (b) PMMA_bP_B, (c) PMMA_bP_C, and (d) PMMA_bP_A.

More in detail, the sample PMMA_bP_B showed intense narrow peaks, as it can be seen by comparing the signal at 16.90° with the broad band associated to PMMA. Even if from the experimental evidences collected by $^{31}$P-MAS NMR analysis the preparation methodology here used (METHOD B) was probably not able to well preserve the bP structure (most of bP seemed to be oxidized), the "surviving" flakes maintained their crystallinity and orientation. The presence of large aggregates was, indeed, also proved by micro Raman analysis, confirming the poor effectiveness of the method in promoting an extensive exfoliation of bP. In other words, these results confirm the poor exfoliation degree of this sample. The sample PMMA_bP_A showed less intense peaks which resembled those already observed for bPn[47,48] even if the content of non-oxidized bP with respect to the degraded portions, as evaluated by NMR, was similar to that of sample PMMA_bP_B. This result implies that METHOD A, starting from suspension of bPn, provided composites with more exfoliated morphology, as suggested also by Raman results. In addition, by repeating the XRD analysis after 6 months (Figure S5), we obtained a completely superimposable curve, assessing that once embedded into the PMMA, the bPn with a certain order degree are stable and the polymer is able to preserve their structure[22].

The sample PMMA_bP_C showed narrow peaks more intense than those of PMMA_bP_A but less intense than those of PMMA_bP_B. On the basis of its highly preserved bP content, assessed by SSNMR (the oxidized/degraded fraction is less than 1/3 of those of hybrids obtained by METHODS A and B), the XRD profile suggested nice output in terms of bP dispersion level and suitability of the METHOD C which can be stated to boost the bP exfoliation and at the same time to preserve the chemical nature and structure of bP nanoflakes.

The stability upon exposure to air and light was also tested by repeating the Raman analysis after 6-10 months from sample preparation and even after prolonged solubilisation of the sample in anisole. Raman spectra collected on different portions of each specimen (films obtained by compression moulding or solution casting) clearly showed the characteristic modes of bP whose intensity and Raman shifts depended on the thickness of flakes: both thicker flakes and polymer portions without visible inclusions evidenced the bP peaks confirming that also thinner flakes (bPn) were not fully etched nor chemically modified after long exposure of the hybrid to air and light (Figure S6).

To better assess the bPn stability in ambient conditions, the photo-degradation induced by UV light irradiation of samples produced by METHOD C was qualitatively studied. The polymer films of PMMA_C_blank and PMMA_bP_C samples were irradiated in air at room temperature with an UV-Vis lamp at different times and the resulting sample analyzed by Raman and FTIR-ATR spectroscopies (Figures 7 and S7, respectively) following a similar approach recently reported[22,49]. The UV light (280 nm) was proved to cause the maximum degradation of mechanically-exfoliated bP flakes (of 20-30 nm thickness), followed by blue light, owing to generation of reactive oxygen species (ROS) participating in bP photo-oxidation[49]. The formation of such species and the role of environmental factors on the photo-oxidation extent were already discussed in the literature, by proposing the bP nanoflakes degradation mechanism and the use of imidazonium salts as effective ROS quenchers [50-53].

Raman spectra of PMMA_bP_C hybrid were recorded by visually heading towards a clean part of specimen (apparently without aggregates, curves labelled (b) in Figure 7) and towards a flake (curve (a) in Figure 7). After 250 min of exposure the ATR spectra of both samples (blank run and hybrid) showed the characteristic vibration modes due to oxidation and degradation effects (Figure S7), *i.e.* absorptions in the region of OH, significant broadening of C=O stretching (see the inset in Fig. S7) and loss of sharpness in the region of fingerprints due to multiple absorptions of oxidized species. In addition, a clear yellow toning for PMMA_C_blank confirmed the degradation effects (see films images before and after UV irradiation on the right of Figure S7).

The Raman analysis (Figure 7) during time of exposure revealed a general loss of the spectra resolution as consequence of polymer degradation, even after the hybrid underwent a 250 min UV irradiation, the signals due to presence of bPn (thinner flakes, curves (b)) were clearly observed, confirming the great stability of nanoflakes once incorporated in the PMMA[22, 49].

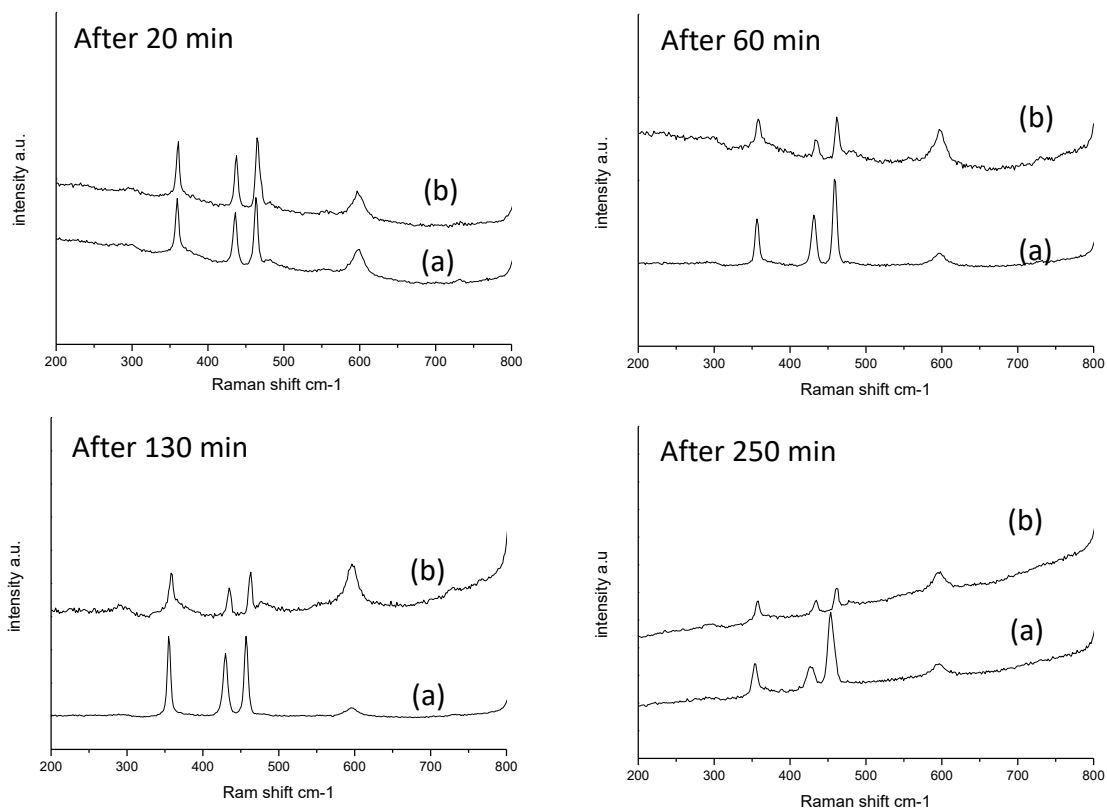

**Figure 7.** Raman spectra of PMMA_bP_C at different time of UV exposure (enlargements in the region of bP modes); curves labelled (a) are referred to visible flakes or aggregates; curve labelled (b) are referred to free/clean portions (without inclusions) of film.

These results confirm the effectiveness of the synthetic approach in preserving the bP nanoflakes structure. It is well-known from literature [50-52] that bP damage is caused by ROS generated by UV light in the presence of oxygen. The polymer chains (PMMA) embedding the nanoflakes protected bPn from oxidation. This was demonstrated by the fact that the UV irradiation of hybrid material provoked the oxidation of PMMA (as clearly shown by ATR), but it had no effect on the bP (as shown by Raman), even by considering the signals attributed to thin flakes; thus we can reasonably conclude that the photogenerated ROS were not able to access the bP surfaces owing to the PMMA sequestration.

**PS- and PNVP- based hybrid materials**

In summary, the *in situ* radical polymerization after LPE of bP in MMA (METHOD C) was here shown as an effective method to provide hybrid PMMA-based materials containing bPn (by promoting the exfoliation of bP) and whose structure was preserved i) during the preparation steps, ii) after storage in ambient condition for prolonged time, iii) owing to different thermal and solvent treatment, and iv) even when subjected to UV aging. In addition, the procedure is simple if compared with other methods and does not involve extensive use of solvents and sonication. To test the feasibility of the method and the possibility to prepare hybrids with different polymers, the *in situ* radical polymerization of Sty and NVP was carried out (Table 1). Two different contents of bP were used for these runs: the hybrids were characterized by SEC, DSC, TGA, Raman and FTIR-ATR spectroscopies and the results were compared to those of their blank experiments (Table 3). In the case of PS hybrids a bimodal shape of MW distribution curves was observed together with certain increase of the $\overline{M_w}$ values, presumably due to confinement of growing macroradicals, as previously discussed for similar PMMA-based sample. These observations suggest that also Sty can establish interactions with bP layers and such interactions are effective for bP exfoliation. In addition, no significant variation concerning the thermal features of all hybrids was observed with respect to blank experiments with the exception of the $T_g$ values which seemed to weakly increase depending on the bP content.

**Table 3:** Molecular weight evolution and thermal features of PS and PVP-based samples

| Sample | $\overline{M_n}$ (D) | $\overline{M_w}$ (D) | $T_g$ (°C) | $T_{onset}$(°C)* | $T_{infl}$(°C)§ |
|---|---|---|---|---|---|
| PS_C_blank | 25,700 | 56,700 | 100.7 | 374 | 418 |
| PS_bP_C | 24,600 | 85,200 | 99.8 | 380 | 417 |
| PS_bP_C2 | 21,000 | 68,100 | 100.2 | 375 | 414 |
| PNVP_C_blank | nd | nd | 160.0 | 410 | 440 |
| PNVP_bP_C | nd | nd | 166.3 | 410 | 440 |
| PNVP_bP_C2 | nd | nd | 163.2 | 412 | 437 |

\* intercept of tangents; § from DTG curves as maximum of peak; nd: not determined

Both PS and PNVP-based materials were analyzed as films obtained by compression moulding. FT-IR spectra showed the vibration modes characteristic of the polymer matrices, whereas the Raman spectra evidenced the typical signature peaks of bP in addition to those of polymers (FTIR and Raman spectra and related attributions are reported in Figure S8 and Table S2 and S3)[54,55]. Moreover, the Raman signals shape and shifts were in agreement with results previously discussed for PMMA-based hybrids. These data confirmed that the synthetic procedure is able to preserve the bP structure, even when different monomers are employed, and to obtain systems potentially suitable for designing devices, which are generally provided by more complex synthetic procedures[56].

After storing the samples in ambient conditions for 6 months, the hybrids PS_bP_C and PVP_bP_C were solubilized in anisole and water, respectively, and films were obtained by solution casting onto glass. They were carefully analyzed by Raman microscopy (Figure 8). Both polymers protected the bP flakes, and the Raman spectra collected in the different parts of the specimen showed the characteristic vibration modes of phosphorus flakes, visible everywhere, although with a different relative intensity. A fine morphology with a good distribution of the particles was found especially for the sample PVP_bP_C, which was obtained by water casting. The microscopic images showed in this case only small and homogeneously dispersed inclusions. Interestingly, for all the inclusions, the Raman spectrum revealed the diagnostic peaks of bP even though the sample was treated with water without protection from air and light. This result definitely underlines the feasibility and the power of the method in providing bPn stabilized as it is prepared and even after different manipulation.

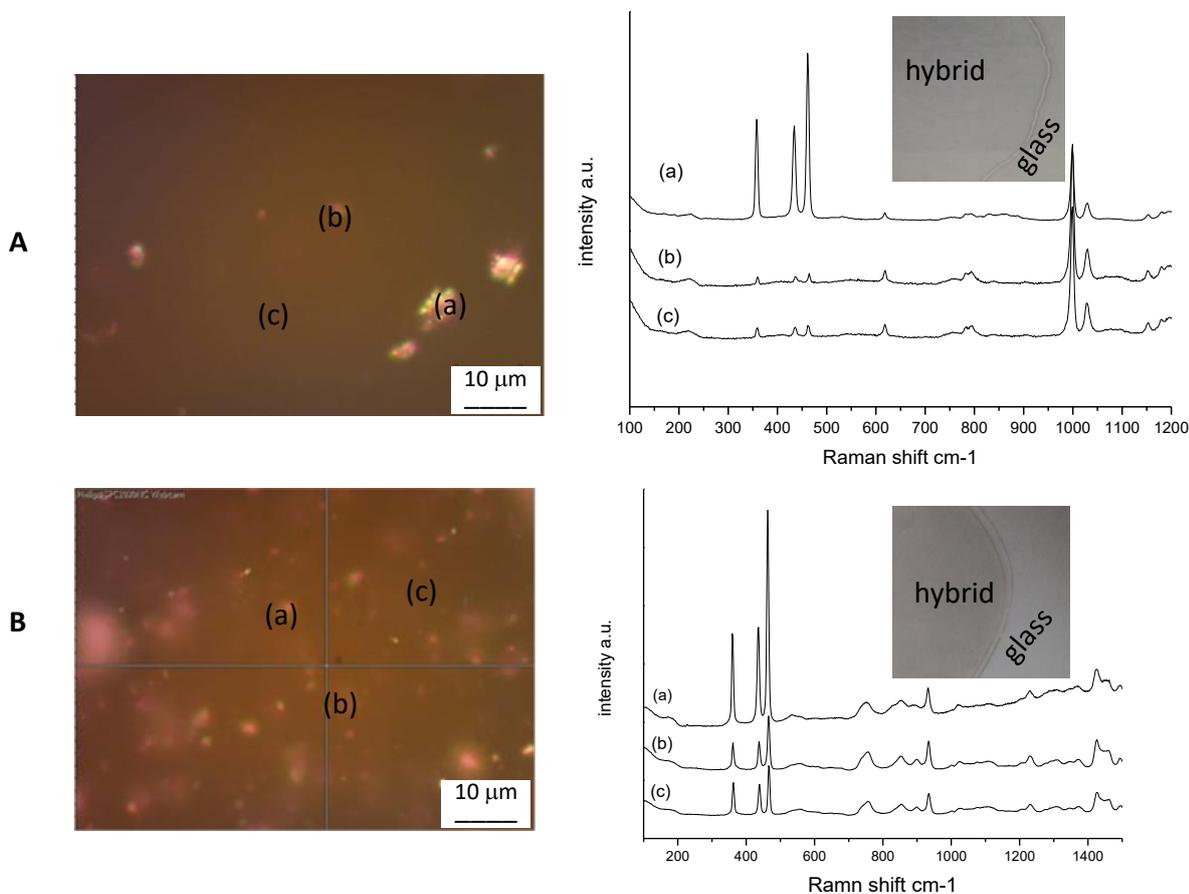

**Figure 8**. Visual imaging and Raman spectra of A: PS_bP_C, and B: PNVP_bP_C, collected at different points of the specimen; Insets: pictures of films obtained onto glasses by solution casting from anisole for hybrid PS_bP_C and from water for PNVP_bP_C.

**Conclusions**

Hybrid materials were obtained by dispersing black phosphorus nanoflakes in polymer matrices through different synthetic strategies with the aim of promoting the exfoliation of bP while protecting the generated nanostructures from oxidation. All the composites ensued processable and their films, obtained by compression moulding, were analyzed by FTIR, Raman, $^{31}$P-SSNMR, XRD, SEC, TGA, DSC to pinpoint the structural characteristics of both the phases: the polymer matrix and the bPn. In addition a deep investigation about the bPn stability, upon different treatments of prepared films (melt processing, solvent solubilisation and UV light irradiation), was performed. Raman, $^{31}$P-SSNMR, XRD analyses evidenced that, depending on the preparation methodology, the hybrids were characterized by a different exfoliation degree and by a different content of bP oxidized species. The procedure comprising the Liquid Phase Exfoliation (LPE) in the vinyl monomer followed by *in situ* radical polymerization provided hybrid polymer-based materials with good dispersion of bP (particularly by using MMA and NVP) and protected bP nanoflakes.

The monomer LPE seems capable of promoting the exfoliation of bP and the following *in situ* polymerization encapsulates the nanoflakes, preserving their structure. By taking into account that, to-date, bPn are prepared in low quantities by mechanical exfoliation, this strategy seems a promising tool to easily provide larger amounts of exfoliated bP. In addition, by considering that the nanoflakes cannot survive for long time in air, light and humidity and that, once generated, bPn have to be passivated by polymer coating, this approach emerges as a new strategy to provide bPn already protected by enveloping the native nanostructures with polymer chains. Therefore, the methodology here realized is able to preserve the bPn structure not only from air and light exposure, but even from thermal and solvent treatment

This approach affords the opportunity to obtain scalable quantities of bPn opening the way for an easier design of (optoelectronic) devices. Moreover, since PMMA is the most used resist for electron beam lithography, solutions of the PMMA nanocomposites can be directly spin-coated without further processing, and the bPn within can be processed into devices without the need of a protective environment for fabrication[57].

**Electronic Supporting Information**

Additional research data supporting this publication, including FTIR-ATR and Raman spectra of starting bP and prepared hybrids, Raman spectra of PMMA_bP_C hybrid after 6-10 months from preparation, FTIR-ATR spectra of PMMA_bP_C hybrid and related polymer matrix after UV aging, FT-IR and Raman peaks assignments of all polymers used, schematic description of synthetic methodologies for hybrids preparation, are available as supplementary file (pdf).

**Acknowledgments**

The European Research Council (ERC) and the National Research Council of Italy (CNR) are acknowledged for funding the work through the project PHOSFUN, an ERC Advanced Grant (Grant Agreement No. 670173), and the project "Ma.Po.Fun" (DCM.AD002.239).

**TOC**

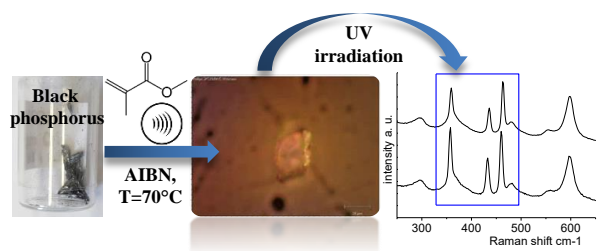